\documentclass[lettersize,journal]{IEEEtran}
\usepackage{titletoc}
\usepackage{diagbox}
\usepackage{algorithmic}
\usepackage{makecell}
\usepackage{algorithm}
\usepackage{array}
\usepackage{amsmath,amssymb,amsfonts}
\usepackage[caption=false,font=normalsize,labelfont=sf,textfont=sf]{subfig}
\usepackage{cite}
\usepackage{textcomp}
\usepackage{stfloats}
\usepackage{url}
\usepackage{verbatim}
\usepackage{graphicx}
\usepackage{lipsum}
\usepackage{xcolor}
\usepackage{makecell}
\usepackage{multirow}
\usepackage{amsmath}
\usepackage{color}
\usepackage[colorlinks,
            linkcolor=blue,
            anchorcolor=blue,
            citecolor=blue,
            urlcolor=blue
            ]{hyperref}

\newtheorem{remark}{\bf Remark}

\begin{document}

\title{On Physics-Informed Neural Network Control for Power Electronics  \\
}
\author{
        \vskip 1em
        Peifeng Hui, \emph{Student Member, IEEE},
        Chenggang Cui, \emph{Member, IEEE},
        Pengfeng Lin, \emph{Member, IEEE},
        Amer M. Y. M. Ghias, \emph{Senior Member, IEEE},
        Xitong Niu\emph{}
       and Chuanlin Zhang, \emph{Senior Member, IEEE}
       
}
\maketitle

\definecolor{limegreen}{rgb}{0.2, 0.8, 0.2}
\definecolor{forestgreen}{rgb}{0.13, 0.55, 0.13}
\definecolor{greenhtml}{rgb}{0.0, 0.5, 0.0}

\begin{abstract}
Considering the growing necessity for precise modeling of power electronics amidst operational and environmental uncertainties, this paper introduces an innovative methodology that ingeniously combines model-driven and data-driven approaches to enhance the stability of power electronics interacting with grid-forming microgrids. 
By employing the physics-informed neural network (PINN) as a foundation, this strategy merges robust data-fitting capabilities with fundamental physical principles, thereby constructing an accurate system model. By this means, it significantly enhances the ability to understand and replicate the dynamics of power electronics systems under complex working conditions. Moreover, by incorporating advanced learning-based control methods, the proposed method is enabled to make precise predictions and implement the satisfactory control laws even under serious uncertain conditions. Experimental validation demonstrates the effectiveness and robustness of the proposed approach, highlighting its substantial potential in addressing prevalent uncertainties in controlling modern power electronics systems.
\end{abstract}


\section{Introduction}
\IEEEPARstart {T}{he} world’s increasing focus on renewable energy resources (RERs) suggests that by 2050, RERs may play a pivotal role in the power systems \cite{b2}. In this context, a vast array of power electronics is being progressively incorporated as interfaces between RERs and loads\cite{b42}. Such integration may lead to diminished system inertia, introducing stability concerns within power systems to varying degrees \cite{b19},\cite{b3}.
The stability issue is largely dependent on power electronics interfaces for voltage regulation. Normally, it is fundamentally to have an accurate modeling process that provide fine control strategies for effective management \cite{b5}. However, when feeding strict control, the power electronics interfaces exhibiting constant power load (CPL) behaviors with negative impedance could potentially compromise system damping and risk overall stability in certain conditions \cite{b5},\cite{b4}. Additionally, the reliability of analytical methods in addressing these stability challenges somewhat depends on the accuracy and accessibility of models and parameters. Such complexity is further aggravated by confidentiality concerns from manufacturers and variabilities due to aging and environmental influences \cite{b18}. These challenges may complicate traditional modeling processes based on large-signal or small-signal analyses\cite{b47}, highlighting the necessity for advanced solutions to ensure the stability of power electronics systems.

By referring to the existing research works, model-driven methods generally rely on the pre-established system models. The employment of nonlinear disturbance observers to monitor CPL changes, employing back-stepping techniques for stabilization, is highlighted in \cite{b7}. Employing virtual impedance technology in source-side converters effectively reduces the output impedance of LC filters, thereby aligning with system stability requirements is illustrated in\cite{b6}. To address the variations in inductance, researchers have developed intuitive observers for inductance, thereby enhancing the robustness of control systems \cite{b23}. Additionally, the introduction of online inductance estimation methods has enhanced the precision of model predictive control (MPC) approaches, with successful applications in dual-active-bridge (DAB) converters \cite{b25}. While model-driven control is highly effective when system dynamics are well understood and accurately modeled, its efficiency diminishes in complex systems where achieving precise modeling is quite a challenging or even impossible.

Data-driven methodologies formulate control strategies by leveraging system data. Research applies MPC for optimization in power electronics, leading to the creation of fully data-driven neural network controllers through imitation learning \cite{b26},\cite{b28}. Although these methods demonstrate notable performance, they often require extensive and diverse training datasets. In situations where data is scarce, such requirements may limit their adaptability \cite{b26},\cite{b28}. Besides, model-free approaches like reinforcement learning (RL) utilize environmental feedback for decision-making, enhancing the control accuracy and promoting the system stability\cite{b30},\cite{b31}. Although these methods are proved to be effective in complex scenarios beyond the scope of conventional modeling, they may encounter challenges related to robustness and interpretability \cite{b33}.

Hybrid approaches exhibit increased robustness in managing complex models by integrating the advantages of both traditional and contemporary control strategies. Mechanisms like learning-based MPC with Gaussian processes (GP) significantly enhance the control precision and adaptability to the dynamic system variations \cite{b35},\cite{b36}. Furthermore, real-time data monitoring and adaptive control via deep neural network (DNN) processors, highlight the hybrid approaches' adaptability and performance efficiency\cite{b38}. These methods offer promising prospects for application. However, their practical implementation, especially in complex scenarios, may encounter challenges associated with computational requirements, data storage needs, and hardware specifications\cite{b28}.

Reflecting on the limitations of traditional model-based controllers and the partial solutions offered by enhanced data-driven methods, this paper is devoted to the integration of artificial intelligence to build a controller with widespread applicability and adaptability. Attention is drawn to Physics-Informed Neural Network (PINN), a powerful modeling tool for complex systems \cite{b40}. PINN merges physics-based model precision with neural network adaptability and learning, requiring minimal initial model specifics. It is skilled at using physical laws to predict dynamics of states not directly measured, thus effectively modeling systems with inherent uncertainties \cite{b12}. In \cite{b13}, PINN extends to address optimal control issues governed by partial differential equations, optimizing loss function weights to solve PDEs. In \cite{b12}, researchers utilize PINN to develop a hybrid modeling framework that combines data-driven insights with physics-based models. This approach enhances model fidelity and outperforms traditional control schemes in generalization and application. Additionally, \cite{b39} explores advanced methodologies for parameter estimation within power electronics, leveraging PINNs to tackle the challenges that conventional models may not adequately address. While PINN behaves promising capabilities, their application in controlling power electronics is still relatively unexplored, highlighting a significant area for future research and practical applications.

\vspace{-1pt}

Aiming to address the complexities and uncertainties in power electronics with significant integration of RERs, this study introduces a fusion of model-driven and data-driven control strategies. By utilizing PINN as a crucial facilitator. the approach enhances the accuracy and flexibility of system modeling by integrating the physical and control laws into the learning process. Furthermore, it ensures robustness and stability through the strategic incorporation of imitative learning techniques. The key achievements of this methodology include: 

\begin{enumerate}
\vspace{-1pt}
\item
Fusion Control Architecture: A PINN-based control framework is introduced, enhancing efficiency in systems integrated with RERs by minimizing dependence on vast datasets and computational power while ensuring system adaptability and robustness.
\item 
Advanced System Modeling: By integrating model-driven and data-driven techniques, the study establishes a highly accurate model for power electronics. This advancement significantly improves the convenience and effectiveness of control methods.
\item
Effective Uncertainty Resolution: The methodology excels in accurately identifying and managing unknown load disturbances and internal circuit parameters, ensuring significantly control performance improvement.
\end{enumerate}
\vspace{-1pt}

Experimental evidence validates the methodology's effectiveness and resilience, demonstrating its substantial potential in overcoming partial integration challenges of RERs in contemporary power electronic systems.
\section{PRELIMINARIES}

\subsection{Power Electronic Systems with Uncertainties}\label{I}

A control system with uncertainties could be conceptualized as a black box model. The input-output relationship of this system is mathematically represented as:
\begin{align}\label{continus dynamic}
    x_{k+1} = f(x_{k}, u_{k}, \theta_{k}, d_{k})
\end{align}
where \( x_k \) signifies the state inferred from current data and \( x_{k+1} \) is the subsequent state. The term \( u_k \) denotes the control input, \( \theta_{k} \) encompasses all possible parameter configurations, and \( d_{k}\) captures model uncertainties and variations. The function \( f \) is designed to approximate system dynamics, mapping the present state and control input to the ensuing state.

The model's parametric nature necessitates a controller design capable of adapting to variations in parameters and disturbances. The existence of uncertainties presents significant challenges to both large-signal and small-signal modeling methods. In light of these challenges, this paper presents an advanced approach that combines model-driven and model-free techniques.

\subsection{Physics-Informed Machine Learning}

\begin{figure}[htbp]
\vspace{-1.0em}
\centerline{\includegraphics[width=0.6\linewidth]{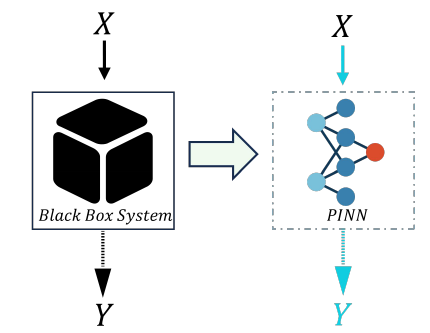}}
\captionsetup{justification=raggedright,singlelinecheck=false}
\caption{The process of fitting a black-box converter model using PINN}
\label{BlackBoxtoPINN}
\vspace{-0.8em} 
\end{figure}

Consider a dynamic system described by
\begin{align}\label{SystemFunction}
    Y = f(X)
\end{align}
where \( X \) and \( Y \) represent the system's state and output, respectively. The PINN fitting process is shown in Fig. \ref{BlackBoxtoPINN}.

\subsubsection{Constructing Hybrid-Physics-Data Models} 
In the construction of Hybrid-Physics-Data Models, a fusion of physics-based numerical models with neural network techniques is employed to enhance the precision and reliability of predictive models.

Initially, a physics-based numerical model $f_{phy} : X \rightarrow Y$ is established, leveraging input variables \ $X$  (which are physically related to the target variable $Y$ ) to simulate the value of $Y$.
Subsequently, a neural network model $f_{nn} : X \rightarrow Y$ is developed, designed to predict by discerning the relationship between the inputs $X$ and the target $Y$.

\subsubsection{Hybrid Physics and Data-driven Loss Function} 
The core design of this loss function is to ensure that the neural network's predictions are consistent with the training data and conform to known physical laws. Such a loss function can be formalized as:
\begin{align}
    \text{arg min}_f \, \text{Loss}(\hat{Y}, Y) + \lambda_{phy} \text{Loss}_{phy}(\tilde {Y}, Y)
\end{align}
where \( \text{Loss}(\hat{Y}, Y) \) represents the discrepancy between model predictions and actual data, \( \text{Loss}_{phy}(\tilde {Y}) \) represents the physical consistency error, and the hyper parameter \( \lambda_{phy} \) controls the weight of physical consistency in training.

\section{PINN Controller Design}\label{II}
The PINN controller design integrates model-driven, data-driven, and machine learning approaches, effectively addressing the challenges posed by integrating renewable energy into power electronics. This framework combines data analytics, physical insights, and control strategies, all fine-tuned through targeted loss functions to ensure adherence to physical and control principles.
\subsection{Framework Design}
At its core, the PINN controller framework employs a parallel architecture encompassing data-driven, physics, control, and loss function modules. This design significantly enhances predictive accuracy and control efficiency.

\subsubsection{Data-driven Module} Utilizes deep learning to forecast system states, disturbances, and parameter variations, thereby improving 
the control precision and resilience.

\subsubsection{Physics Module} Integrates physical laws into neural network forecasts, ensuring predictions are both data-driven and physically consistent. This approach enhances understanding of system behavior and improves the accuracy of complex system predictions.

\subsubsection{Control Module}Relies on outputs from the data-driven and physics modules to dynamically adjust control laws, ensuring robust and adaptive responses to system changes and disturbances for maintained stability under varied conditions.

\subsubsection{Loss Function Module}
Incorporates physics-based and control-based loss functions to refine neural network training, aligning learning outcomes with physical laws and control objectives for optimal system performance.

\begin{figure}[htbp]
\vspace{-1.0em}
\centerline{\includegraphics[width=1.0\linewidth]{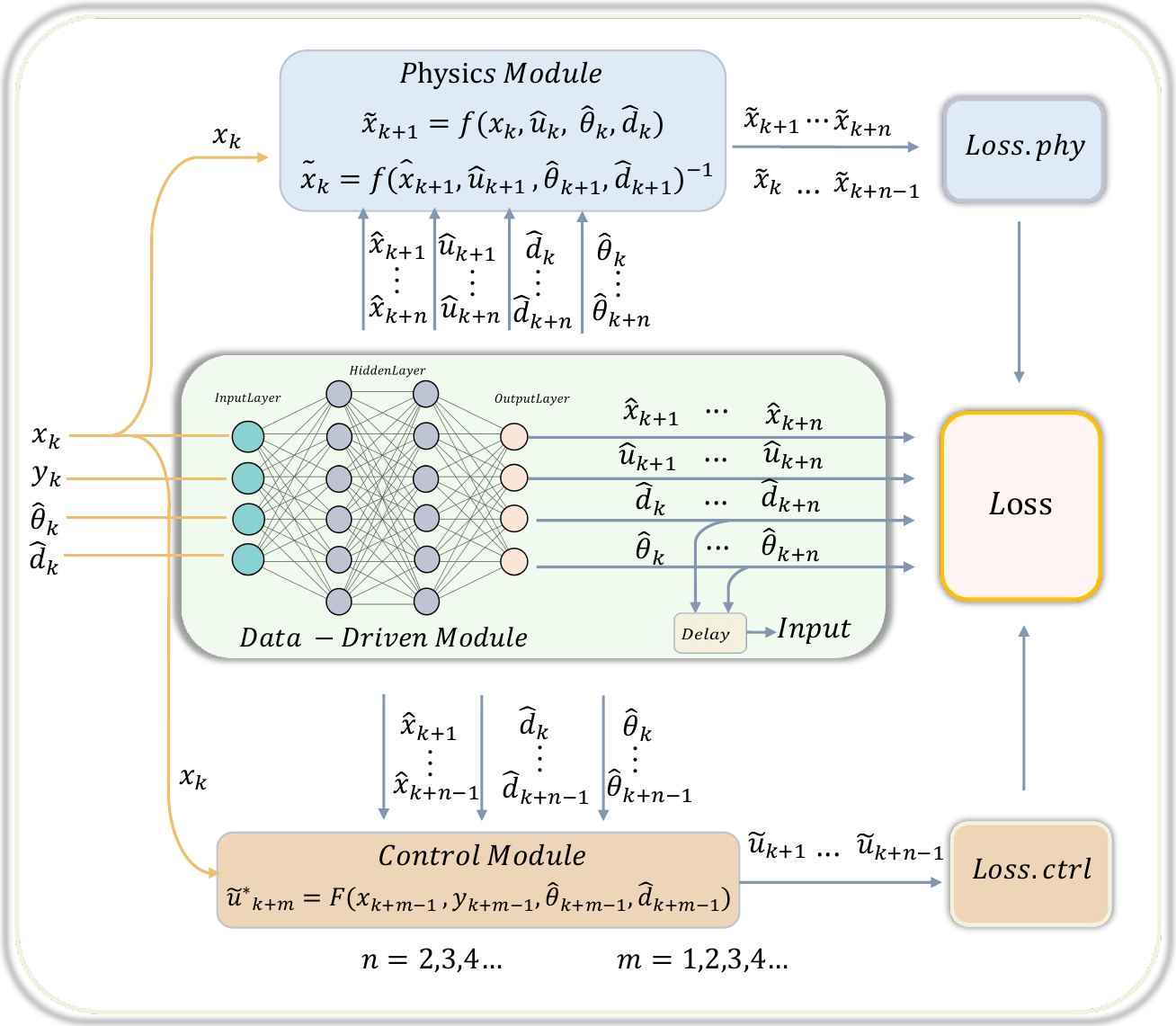}}
\caption{Structure of the proposed hybrid intelligent control framework and its training process.}
\label{PINNsframework}
\vspace{-0.8em} 
\end{figure}

Fig. \ref{PINNsframework} illustrates the PINN control framework and its training process, demonstrating the structured approach to hybrid intelligent control.
where $n$, $x_k$, $y_k$, $\left[ {{{\hat x}_{k+1}},{\rm{\cdot\cdot\cdot,}}{{\hat x}_{k + n}}} \right]$, $\left[ {{{\hat d}_k},{\rm{\cdot\cdot\cdot,}}{{\hat d}_{k + n}}} \right]$, $\left[ {{{\hat \theta }_k},{\rm{\cdot\cdot\cdot,}}{{\hat \theta }_{k + n}}} \right]$, $\left[ {{{\hat u}_k},{\rm{\cdot\cdot\cdot,}}{{\hat u}_{k + n}}} \right]$ represent the prediction horizon in a sequential time series, current selected system state variables, current system output, predicted state variables, disturbances, system parameters and control variable, respectively.  Detailed discussions on ${\tilde u^*}$ and ${{\tilde x}_k}$ will follow, elucidating the control and prediction mechanisms within the framework.

\subsection{Physics Module Design}
The physics module embeds fundamental physical laws into the training process of the framework, allowing for joint optimization and adherence to these principles. A discrete state-space representation, derived from a time-varying parameter system approach, ensures that model predictions are both data-informed and physically consistent:
\begin{align}\label{discrete equation}
{\tilde x_{k + 1}} = {A_k}({\hat \theta _k}){x_k} + {B_k}({\hat \theta _k}){u_k} + {E_k}({\hat \theta _k}){\hat d_k}, \\ {y_k} = {C_k}({\hat \theta _k}){x_k},\quad\quad\quad\quad\quad
\end{align}
where \( A(\theta) \), \( B(\theta) \) and \( E(\theta) \) respectively represent the system, input, and disturbance matrices, all of which vary with parameters with respect to time. \( C(\theta) \) denotes the output matrix. These matrices collectively describe how system states, inputs, and disturbances translate into outputs, capturing the system's physical dynamics. 

To facilitate real-time prediction and system adaptation, the data-driven module forecasts parameters such as $\hat d_{k}$, $\hat \theta_k$, and future states $\hat x_{k+1}$. The backward equation (\ref{backward}), derived for revising predictions based on new data, complements the forward dynamics (\ref{discrete equation}) by enabling an iterative refinement process:
\begin{align}\label{backward}
\begin{aligned}
{{\tilde x}_k} = ({A_{k + 1}}({{\hat \theta }_{k + 1}}){{\hat x}_{k + 1}} + {B_{k + 1}}({{\hat \theta }_{k + 1}}){{\hat u}_{k + 1}}\\
 + {E_{k + 1}}({{\hat \theta }_{k + 1}}){{\hat d}_{k + 1}}{)^{ - 1}}.
\end{aligned}
\end{align}

This iterative prediction and refinement process, underpinned by physical laws, not only enhances the neural network's prediction accuracy but also ensures the results are consistent with physical realities. Significantly boosting the reliability and relevance of predictions in power electronic systems.

\subsection{Control Module Design}
The control module leverages neural networks to internalize the dynamics governed by advanced controllers, thereby facilitating the formulation of autonomous control strategies. This module dynamically computes control variables using parameters predicted by the data-driven module, aiming to optimize the system's cost function for enhanced adaptability and control precision. The system's cost function \( J \) is designed as:
\begin{align}
\begin{split}
J(u) = \sum\limits_{i = 1}^n {(Q{{({y_{k + i}} - {y_{ref}})}^2}}  + R{({u_{k + i}} - {\hat u_{k + i}})^2})
\end{split}
\end{align}
where \( Q \) and \( R \) denoting matrices that weight predictive tracking error and control input, over a prediction horizon $n$ is the prediction horizon. The derivation of the optimal control law, $\tilde u^*$, is formulated as a function of current states, predicted parameters, and disturbances:
\begin{align}\label{optimalcontrollaw}
\tilde u^* = F (x_k,y_k,{\hat \theta _k}, {{\hat d}_k}).
\end{align}

Meanwhile, the cost function can be included in the overall loss function for joint optimization. Incorporating these elements into the neural network facilitates a dynamic adjustment of control strategies to accommodate system changes and disturbances. 
\begin{remark}
Unlike traditional methods that depend heavily on data, the fusion control strategy emphasizes model-based control law learning, reducing the reliance on extensive datasets. This method is adaptable beyond mere imitation learning for MPC, showing potential across a range of control algorithms with enhanced generalization and optimization.
\end{remark}

\subsection{Data-driven Module Design}\label{AA}

For effective implementation of control strategies, it is crucial to accurately identify 
the disturbances and variations in system parameters. This study utilizes a data-driven approach to enhance the effectiveness of control strategies by employing a neural network for predictive modeling. The neural network models future system states \(\hat x_{k+n}\), control laws $\hat u_{k+n}$, estimate system parameters \(\hat \theta_{k+n}\) and disturbances \(\hat d_{k+n}\) for $n=0,1,2,3,...$. This methodology is able to enhance the precision and robustness of control strategies.

\subsubsection{Inputs}
Inputs for the data-driven model include current state variables \(x_k\) and $y_k$. Additionally, estimated disturbances \(\hat d_{k}\) 
and parameters \(\hat \theta_{k}\) derived from previous outputs, are integrated as inputs to refine prediction accuracy and enhance generalization.

\subsubsection{Outputs}
The data-driven component forecasts future states \(x_{k+n}\) and control outputs $\hat u_{k+n}$, alongside estimated parameters \(\theta_{k+n}\) and disturbances \(d_{k+n}\) for $n$=1, 2, 3.... This model extends beyond traditional prediction by estimating uncertainties like parameters or disturbances, using these estimates to improve the physical and control modules' dynamic accuracy and adaptability.

\subsubsection{Hidden Layers Configuration}
The neural network architecture incorporates flexible hidden layers, designed to model the system's complex dynamics efficiently. The configuration of these layers—whether dense, convolutional, or recurrent—is customized to meet the specific requirements of the system. Such adaptability enhances the model's ability to generalize, enabling it to accurately predict future states and improve control strategies effectively.

By integrating empirical data with foundational physical knowledge, the data-driven module serves as a cornerstone for developing an accurate, robust, and adaptive control strategy, underpinning the efficacy of the entire hybrid intelligent control framework. 

\begin{remark}
    It is worth noting that the prediction length $n$  can be flexibly chosen based on specific needs. For configuring data-driven output, one simply needs to add the respective channels of $x_{k+n}$, $u_{k+n}$, and other predicted values based on the selected value of $n$, while configuring the corresponding physics module and control module. This approach exhibits a satisfactory generalization performance.
\end{remark}

\subsection{Physics-based and Control-based Loss Functions Design}
To ensure that models align with physical knowledge, this study implements physics-based and control-based loss functions for training data-driven models. These loss functions are designed to maintain consistency with physical principles and control laws while optimizing the model's predictive accuracy on the training set with minimal complexity. The data-driven model's loss function is formulated as:
\begin{align}
{\rm{Loss}}.{\rm{data}} = \frac{1}{n}\sum\limits_{i=0}^n \begin{split}
(({x_{k + i}},{{\hat x}_{k + i}});({d_{k + i}},{{\hat d}_{k + i}});\\({u_{k + i}},{{\hat u}_{k + i}});({\theta _{k + i}},{{\hat \theta }_{k + i}})).
\end{split}
\end{align}

To ensure predictions are in line with physical and control principles, the following physics-based and control-based loss functions are utilized:

\begin{equation}
{\rm{Loss}}{\rm{.phy}_{forward}} = \frac{1}{n}\sum\limits_{i=1}^n \begin{array}{l}
||({x_{k + i}},{{\tilde x}_{k + i}})|{|} + ||({{\tilde x}_{k + i}},{{\hat x}_{k + i}})|{|},
\end{array} 
\end{equation}
\begin{equation}
{\rm{Loss}}.{\rm{ph}}{{\rm{y}}_{{\rm{backward}}}} = ||({x_k},{\tilde x_k})|| + ||({\tilde x_k},{\hat x_k})||,
\end{equation}
\begin{equation}
\begin{array}{l}
{\rm{Loss}}.{\rm{control}} = \frac{1}{n}\sum\limits_{i = 1}^n {\begin{array}{*{20}{l}}
{(||({u_{k + i}},{{\tilde u}_{k + i}})|{|^2} + ||({{\tilde u}_{k + i}},{{\hat u}_{k + i}})|{|^2})}
\end{array}}\\ + J(u).
\end{array}
\end{equation}

The comprehensive learning objective, incorporating both \(\text{Loss.phy}\) and \text{Loss.control}, is defined as:
\begin{equation}
\begin{array}{l}
{\rm{argmin }}\quad{\lambda _{{\rm{phy}}}}({\rm{Loss}}.{\rm{ph}}{{\rm{y}}_{forward}}{\rm{ + Loss}}.{\rm{ph}}{{\rm{y}}_{backward}})\\
 \quad\quad\quad+ {\rm{Loss}}.{\rm{data}} + {\lambda _{{\rm{control}}}}{\rm{Loss}}.{\rm{control}},
\end{array}
\end{equation}
where \( \lambda_{\text{phy}} \) and \( \lambda_{\text{ctrl}} \) are hyperparameters that balance the importance of minimizing physics inconsistencies and empirical loss against model complexity.

This advanced loss function framework not only focuses on the accuracy of predictions but also emphasizes adherence to physical principles and control law adaptability, setting it apart from traditional loss function approaches.

\section{Application to Power Electronic Systems}
In this section, a series of experimental and simulation studies employs a DC-DC buck converter as a case to explore the effectiveness of the proposed PINN controller.

\begin{figure*}[htbp]
\vspace{-0.8em}
\centerline{\includegraphics[width=1.0\linewidth]{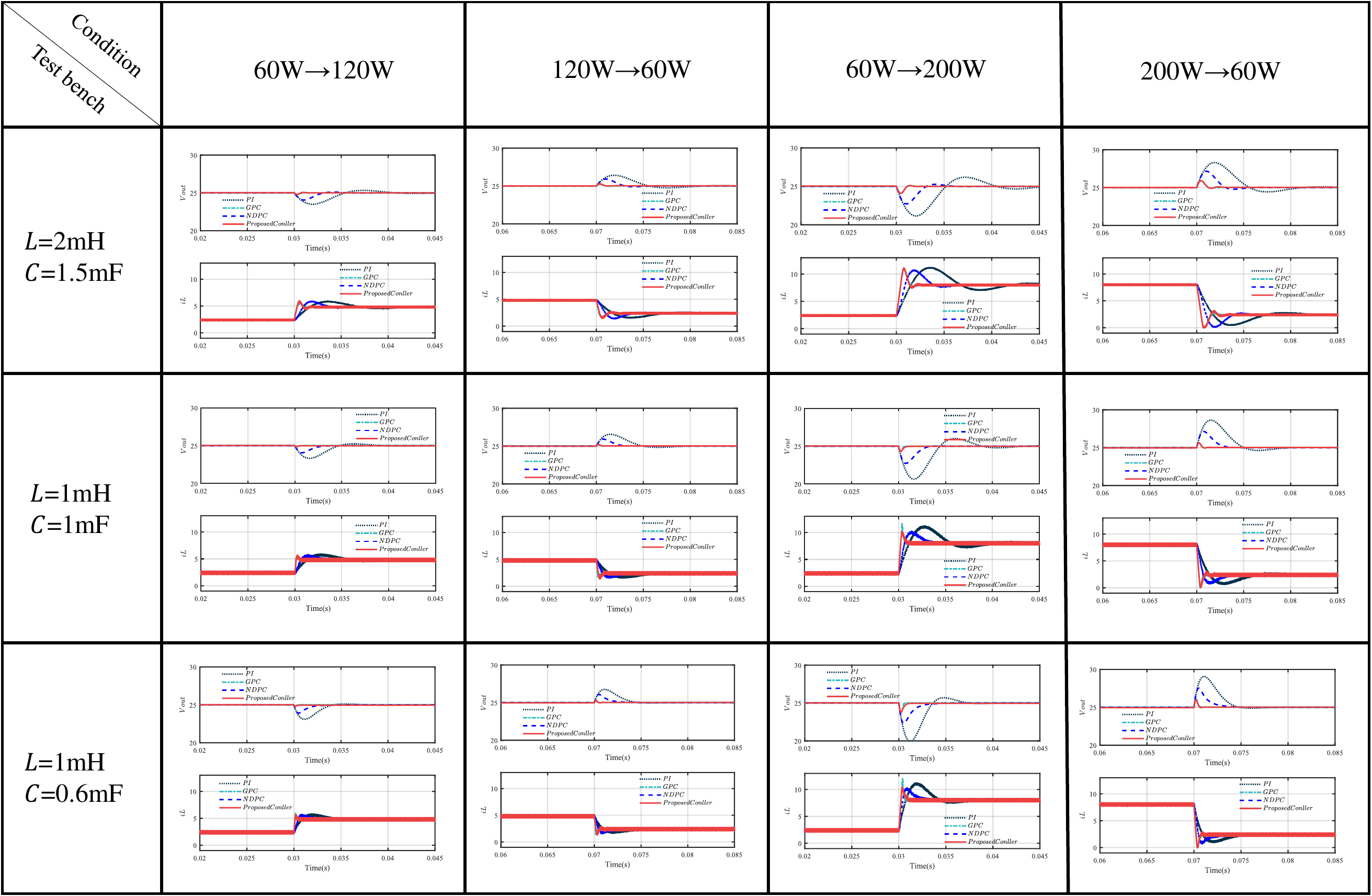}}
\caption{Validation of the proposed control method through adaptability testing and comparison with alternative methods.}
\label{Comparision1}
\vspace{-0.8em} 
\end{figure*}

\subsection{Dynamic Model Descripiton}
The dynamic model of the buck converter can be expressed by the following form\cite{b5}:
\begin{align}\label{dynamicmodel}
\left\{ \begin{array}{l}
\Delta {i_L} =  - \frac{1}{{{L_N}}}{v_o} + \frac{{{V_{in}}}}{{{L_N}}}u + \frac{1}{{{L_N}}}{\eta_1}\\
\Delta {v_o} = \frac{1}{{{C_N}}}{i_L} + \frac{1}{{{C_N}}}{\eta_2}
\end{array} \right.
\end{align}
where ${v_o}$ is the output voltage, $i_L$ is the inductor current, $u$ is the control duty ratio, $V_{in}$ is the Input Voltage of the source bus; ${L_N}$ and ${C_N}$ are the nominal values of inductance and capacitance, respectively; $\eta_1$and $\eta_2$ represent lumped uncertainties. $\eta_1$ is typically represented as 
\begin{align}
    {\eta_1} =  - \Delta L\Delta {i_L}.
\end{align}

When feeding an impedance load $R_{imp}$, $\eta_2$ can be described as:
 \begin{align}
 {\eta_2}_{imp} =  - \frac{{{v_o}}}{R_{imp}} - \Delta C\Delta {v_c}.    
 \end{align}
 
When feeding a constant power load, $d_2$ can be described as:
\begin{align}
 {\eta_2}_{CPL} =  - \frac{{{P_{CPL}}}}{v_o} - \Delta C\Delta {v_c}.    
\end{align}

The control objective is to provide satisfactory reference output voltage tracking in the presence of system uncertainties.

\subsection{Simulation Configuration}\label{simulation}
The proposed controller is validated in Matlab/Simulink simulations to verify its effectiveness. The input set ${S_k}$ is expressed as
\begin{align}
{S_k} = \left\{ \begin{array}{l}
{v_{o\_del_{k}}}, {v_{o_{k}}}, \Delta {v_{o_{k}}}, {i_{L\_del_{k}}}, {i_{L_{k}}}, \Delta {i_{L_{k}}},\\
{y_{\_del_{k}}}, {y_{k}}, \Delta {y_{k}}, {d_{k}}, {C_{k}}, {L_{k}}
\end{array} \right\},
\end{align}
where $y_k$ denotes the voltage tracking error, $d_k$ represents load variation disturbances, and $\theta_k$ signifies real-time capacitance $C_k$ and inductance $L_k$. The set also includes capacitance voltage ${v_{o_{k}}}$, inductor current ${i_{L_{k}}}$, their respective delayed values ${v_{o_{del_{k}}}}$ and ${i_{L_{del_{k}}}}$, along with their variations $\Delta {v_{o_{k}}}$ and $\Delta {i_{L_{k}}}$.
\begin{table}[htbp]
\vspace{-1.0em}
        \centering
        \caption{Simulation and training parameter setting}
        \label{Simulation and training parameter setting}  
        \begin{tabular}{c c c}
                \hline\hline\noalign{\smallskip}        
                Parameters & Description & Value \\
                \noalign{\smallskip}\hline\noalign{\smallskip}
                $V_{ref}$  & Load Bus Reference Voltage  & 25 V{\smallskip} \\
                $f_{s}$   & Switching Frequency & 20 kHz{\smallskip} \\
                $V_{in}$   & Input Voltage  & 50 V{\smallskip} \\
                $M$   & Hidden Layers  & $2 \times 32$ {\smallskip} \\
                $L_N$  & Inductance & 2 mH{\smallskip}\\
                $C_N$& Capacitance & 1mF{\smallskip}\\
                $Q$, $R$& Cost Function weight & 5, 1{\smallskip}\\ 
                $\lambda_{\text{phy}}$, $\lambda_{\text{ctrl}}$& Loss hyperparameters & 0.6, 10{\smallskip}\\ 
                $R$& Case impendence load & 4.037,$\Omega$, 2.537,$\Omega$, 5.759,$\Omega${\smallskip}\\
                $CPL$& Case CPL load & 60W, 150W, 100W{\smallskip}\\
                \hline\hline
        \end{tabular}
\end{table}

For training data acquisition, an offset free generalized predictive controller\cite{b42} is utilized, setting a simulation step size at 1e-6 seconds to align with a sampling frequency of 1 MHz. The data set comprises 18,000 samples, spanning from the $10^{th}$ to the $18009^{th}$ data point, partitioned into $75\%$ for training and $25\%$ for validation purposes. The prediction horizon, $n$, is established at six steps. Training utilizes MATLAB's Deep Learning Toolbox, incorporating network with a custom loss function. The process completes in approximately 45 minutes on a platform equipped with an AMD Ryzen 5 7600 6-Core Processor operating at 3.80 GHz. Additional details on simulation and training parameters are provided in Table \ref{Simulation and training parameter setting}.

The data-driven component's neural network theoretically requires sufficient expressive capability to model complex intermediate states. Therefore, it is necessary to evaluate the network architecture across different configurations of hidden layers and neurons. Tests conduct on various combinations of hidden layer configurations and training iteration numbers, with the results presenting in Table \ref{NN Configurations}. The training results indicate that, as the number of layers and neurons gradually increases, the PINN is able to acquire better expressive power. Besides, Note that when the neuron count exceeds 16 and the number of layers surpasses 2, the average error stabilizes below 5\%, suggesting the network's expressiveness is adequate for this parameter estimation task. Considering computational efficiency and accuracy, a slightly larger network structure, comprising a deep neural network with 2 layers and 32 neurons per layer, is chosen as the final design for the data-driven component.

\begin{table}[htbp]
\centering
\renewcommand{\arraystretch}{1.5} 
\caption{Average Percentage Error (\%) of Different Neurons and Layers}
\label{NN Configurations}
\begin{tabular}{c|c c c c c c}
\hline
\diagbox[dir=SE,width=7em]{Neurons}{Layers.} & \multicolumn{1}{|c|}{$1$} & \multicolumn{1}{c|}{$2$} & \multicolumn{1}{c|}{$3$} & \multicolumn{1}{c|}{$4$} & \multicolumn{1}{c|}{$5$} &{$6$} \\ \hline
 8 & 20.7\% & 16.8\% & 13.4\% & 6.1\% & 4.2\% & 4.2\% \\ 
 16 & 18\% & 4.2\% & 3.2\% & 3.7\% & 2.2\% & 1.9\% \\ 
 32 & 6.6\% & 1.9\% & 2.1\% & 1.8\% & 1.8\% & 1.7\% \\ 
 64 & 5.3\% & 2.0\% & 2.3\% & 1.8\% & 1.9\% & 1.7\% \\  \hline
\end{tabular}
\end{table}

As illustrated in Table \ref{TrainLoss}, it presents the convergence process. With the increase in training iterations, the comprehensive average error (\%) gradually decreases. Ultimately, the training error stabilizes within 1.9\%.

\begin{table}[htbp]
\centering
\renewcommand{\arraystretch}{1.5} 
\caption{ Convergence of the training process and the average
percentage error}
\label{TrainLoss}
\begin{tabular}{c|c c c c c c}
\hline
{Iterations} & \multicolumn{1}{|c|}{$0.1{e^5}$} & \multicolumn{1}{c|}{$0.5{e^5}$} & \multicolumn{1}{c|}{$1.5{e^5}$} & \multicolumn{1}{c|}{$2{e^5}$} & \multicolumn{1}{c|}{$2.5{e^5}$} &{$3{e^5}$} \\ \hline
 {Average Loss} & 13.2 & 6.5 & 4.7 & 5.3 & 3.4 & 2.2 \\ 
 {Percentage Error} & 12.7\% & 7.8\% & 3.9\% & 2.3\% & 2.1\% & 1.9\% \\ \hline
\end{tabular}
\end{table}

\subsection{Simulation Verification}

The simulation results is shown in Fig. \ref{Comparision1}. Considering the fact that the system stability is most challenged when feeding pure CPL, this study employs CPL connected to the load bus for simulation comparisons. Three distinct sets of parameters and two operational conditions are defined, enabling a comprehensive comparison against nonlinear disturbance observartion GPC\cite{b48}, dual PI controller, and the offset free GPC\cite{b42}. Initially, a 60W CPL is connected to the bus. Subsequently, the CPL escalates to 120W and 200W at 0.03s, before reverting to 60W at 0.07s. Observations indicate that the proposed method adeptly acquires control behaviors from the target controller, ensuring voltage stability during load fluctuations, and either matches or exceeds the performance of alternative controllers. Furthermore, the method promptly recognizes changes in system parameters without dependency on the observer of the traditional GPC method for managing uncertain parameters. This approach results in accelerated parameter identification, minimized observer modeling duration, and enhanced dynamic performance in comparison to PI control.

\subsection{Generalization Analyses}
This section elaborates on generalization error analysis applied to the PINN controller design, leveraging VC dimension and generalization bounds concepts\cite{b43, b44, b45}. The generalization error, validation error, and training error are denoted by $E_{gen}$, $R_{test}$ and $R_{emp}$ respectively, with the generalization error expressed as:
\begin{align}\label{generalization error}
{{E_{gen}} = {R_{test}} - {R_{emp}}}.
\end{align}

Given a sample size $m$ and a probability threshold $\varepsilon > 0$, applying probability theory to this equation yields:
\begin{align}
\mathop {\lim }\limits_{m \to \infty } {R_{test}} - {R_{emp}} \ge \varepsilon  = 0.
\end{align}

To address scenarios where the generalization gap exceeds $\varepsilon$, hoeffding's inequality\cite{b46} is utilized to establish a probability bound:
\begin{align}\label{heoffding eq}
\varepsilon  \le \mathop {\lim }\limits_{m \to \infty } {R_{test}} - {R_{emp}} \le 2\left| H \right|\exp ( - 2m{\varepsilon ^2}),
\end{align}
where $|H|$ denotes the model complexity. Using basic algebraic knowledge, denoting $\varepsilon$ by $\delta$, setting a confidence coefficient $(1 - \delta)$ and modifying the above equation results in:
\begin{align}\label{22}
    {{R_{test}} \le {R_{emp}} + \sqrt {\frac{{\ln \left| H \right| + \ln \frac{2}{\delta }}}{{2m}}} },
\end{align}
where $\delta$ representing the failure probability. Fulfilling this criterion indicates the model possesses robust generalization capabilities. For this neural network architecture with 15 input layers, 2x32 hidden layers, and 28 output layers, setting $\delta = 5\%$ and $m$ = 14500, the  model complexity $H$ is calculated as 2492.

Applying (\ref{22}) and defining the hoeffding equation as $\Phi$, where $\Phi=1.99\%$, aligns with the simulation outcomes presented in \ref{simulation}. The validation error ($R_{test}$) is documented at 3.03\%, and the training error ($R_{emp}$) at 1.9\%, indicating 
the fine generalization performance of the PINN controller.

\section{Experimental Results}

\begin{figure}[htbp]
\centerline{\includegraphics[width=1\linewidth]{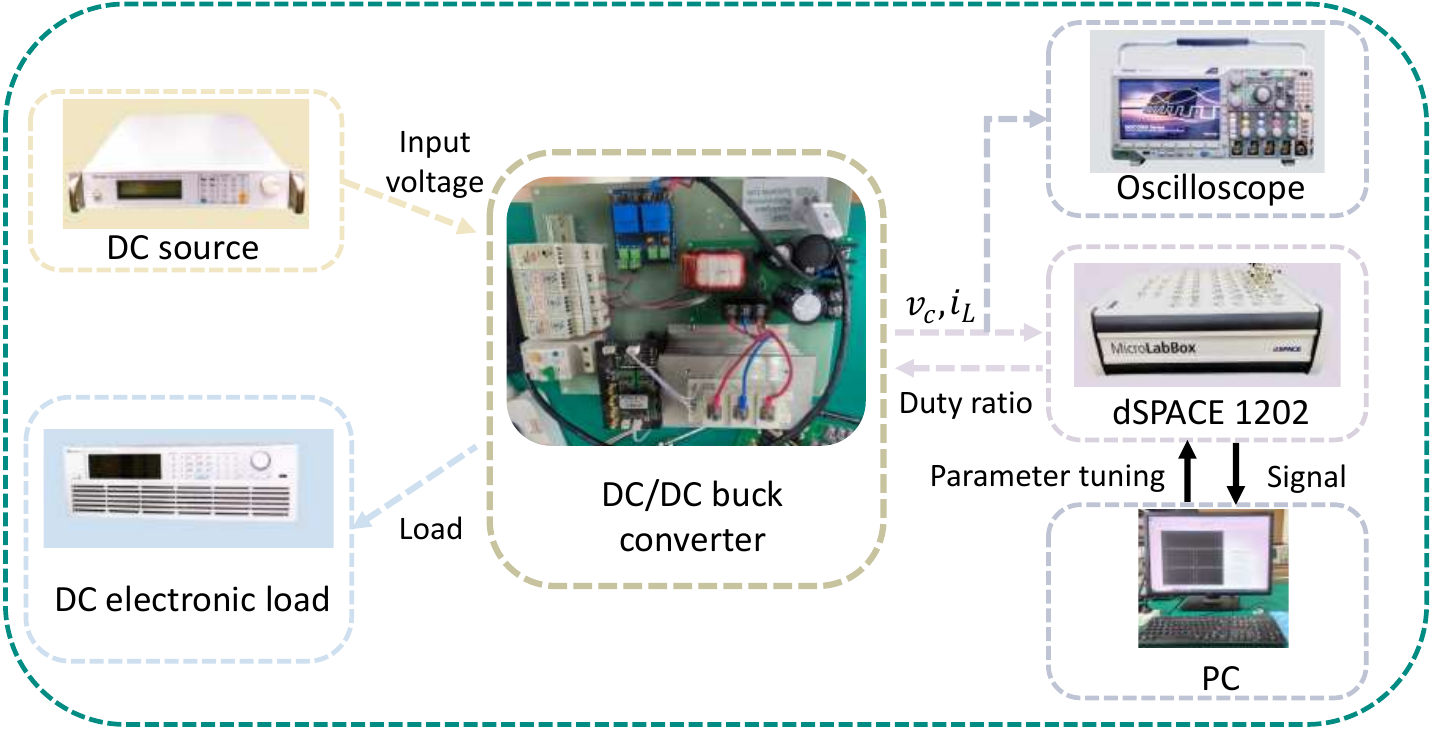}}
\caption{ Experimental setup.}
\label{equipment}
\vspace{-1.0em}
\end{figure}

To validate the effectiveness of the proposed method, an experimental setup was constructed as illustrated in Fig.\ref{equipment}. This setup comprises a programmable DC power supply, an electronic load, a DC/DC buck converter, and a dSPACE 1202 system. The proposed control methodology is executed on the dSPACE platform, which generates PWM signals for the DC/DC buck converter (C=0.9mF, L=1mH). To simulate constant power load (CPL) conditions, the Chroma programmable load is configured to operate in constant power mode.

\begin{figure}[htbp]
\centerline{\includegraphics[width=1\linewidth]{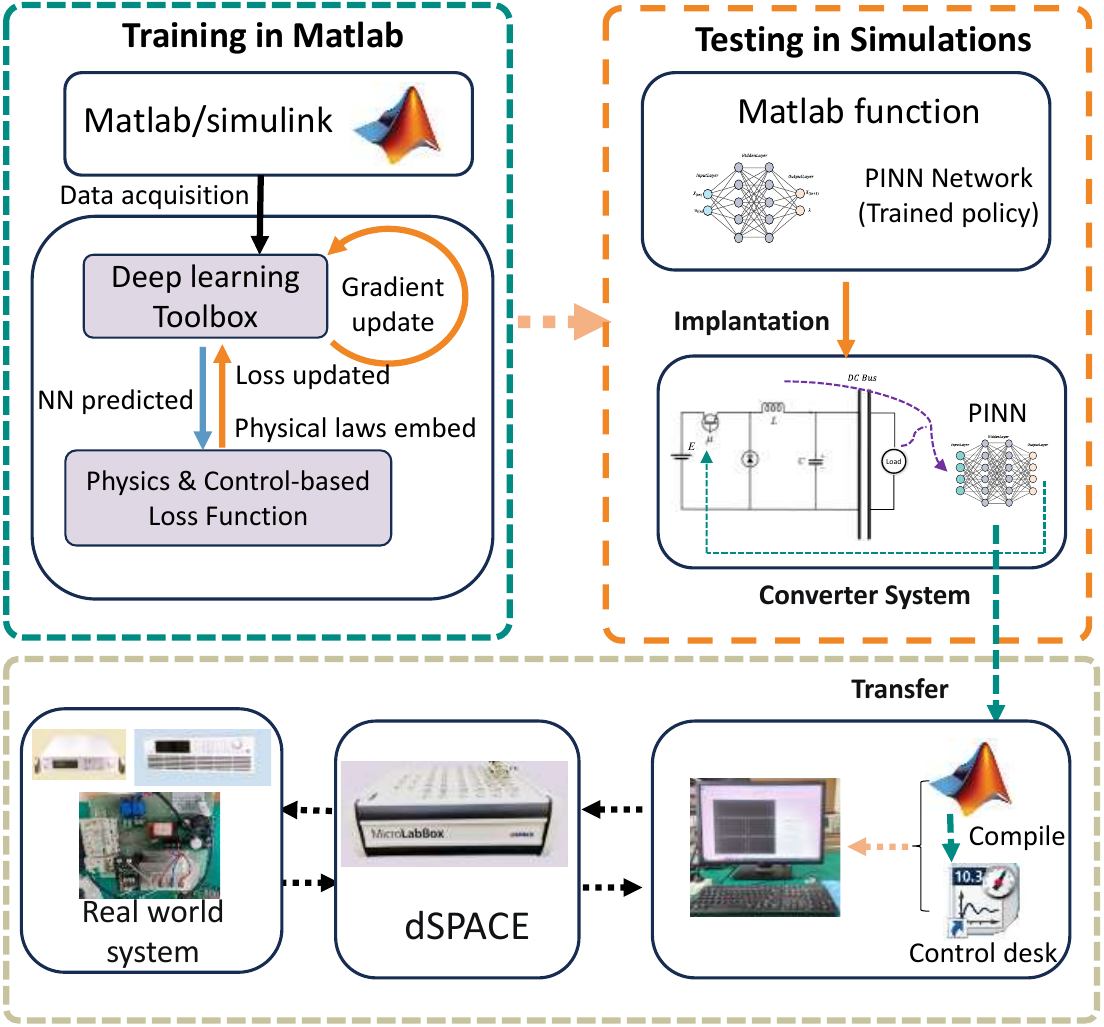}}
\caption{The experiment implementation process of the proposed method.}
\label{TransferProcess}
\vspace{-1.0em}
\end{figure}

The implementation of the proposed method unfolds across three sequential stages, as illustrated in Fig. \ref{TransferProcess}.  Initially, simulations generate data to the deep learning (DL) toolbox, where physics-based and control-based loss functions guide neural network predictions and backpropagation. This process effectively models the power electronics system and cultivates control laws via imitation learning. The second stage involves refining the neural network by adjusting trained weights and biases, creating an enhanced network that embodies the optimal policy. This network, encapsulated within a Matlab function, serves as a dynamic tool for monitoring parameters and refining control strategies using real-world data. In the final stage, the controller is converted into C code through a compiler, enabling its integration with a dSPACE MicroLabBox that operates on a real-time kernel. This allows for the immediate deployment in power electronics systems.

\subsection{Comparison With Various Work Conditions}
\begin{figure}[htbp]
\centerline{\includegraphics[width=1\linewidth]{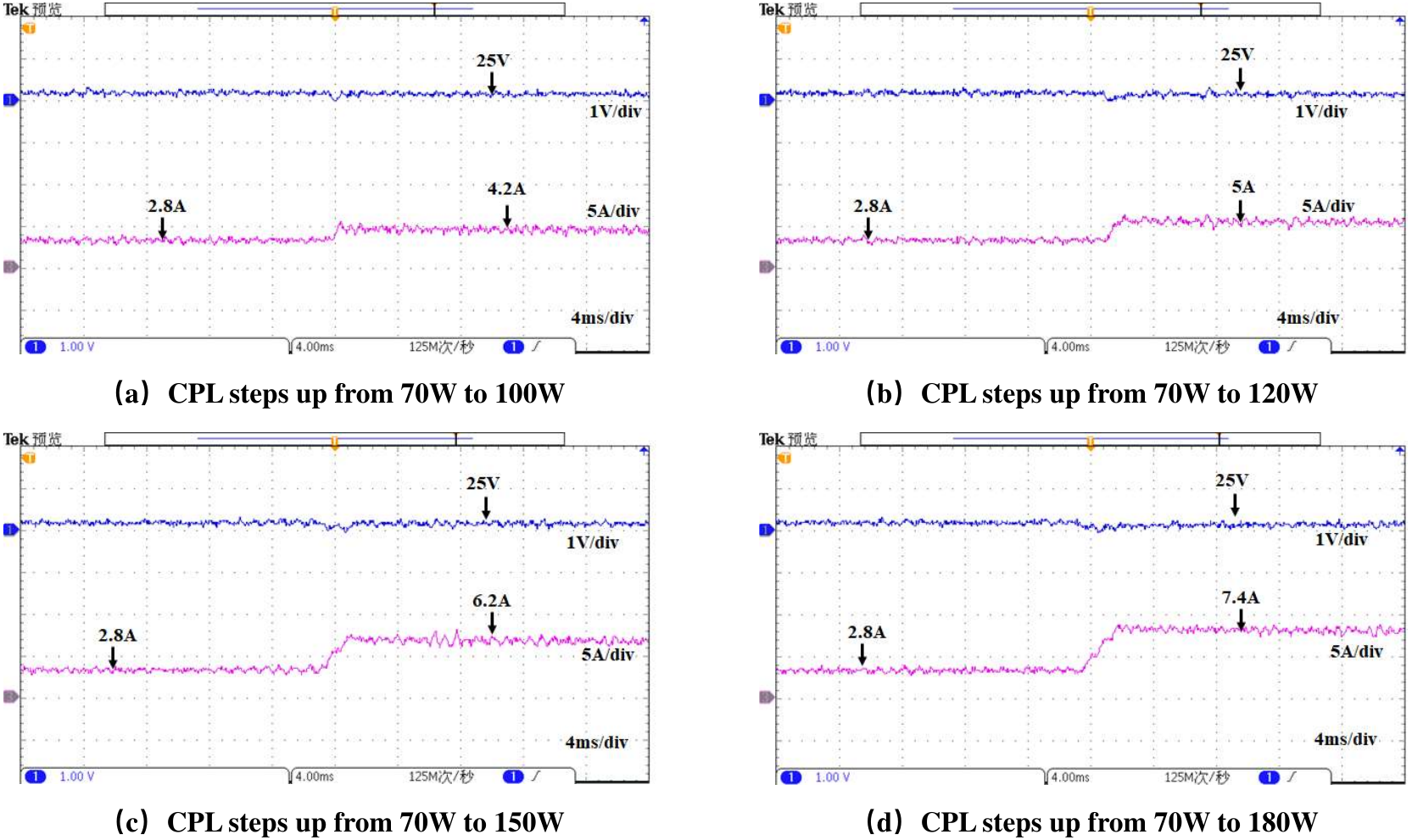}}
\caption{Experimental results of different operating conditions.}
\label{DifferentWorkConditionsComparsion}
\vspace{-1.0em}
\end{figure}

Experimental results depicted in Fig. \ref{DifferentWorkConditionsComparsion} demonstrate the effectiveness of the proposed method under a variety of operating conditions. As the system states evolve, the method automatically adapts to load changes, ensuring superior output voltage tracking performance.

\subsection{ Performance Comparison With Other Controllers}

\begin{itemize}
\item[$\bullet$] Case I. Comparisons analysis under $R_{imp}$ conditions:
\end{itemize}  

\begin{figure}[htbp]
\centerline{\includegraphics[width=1.1\linewidth]{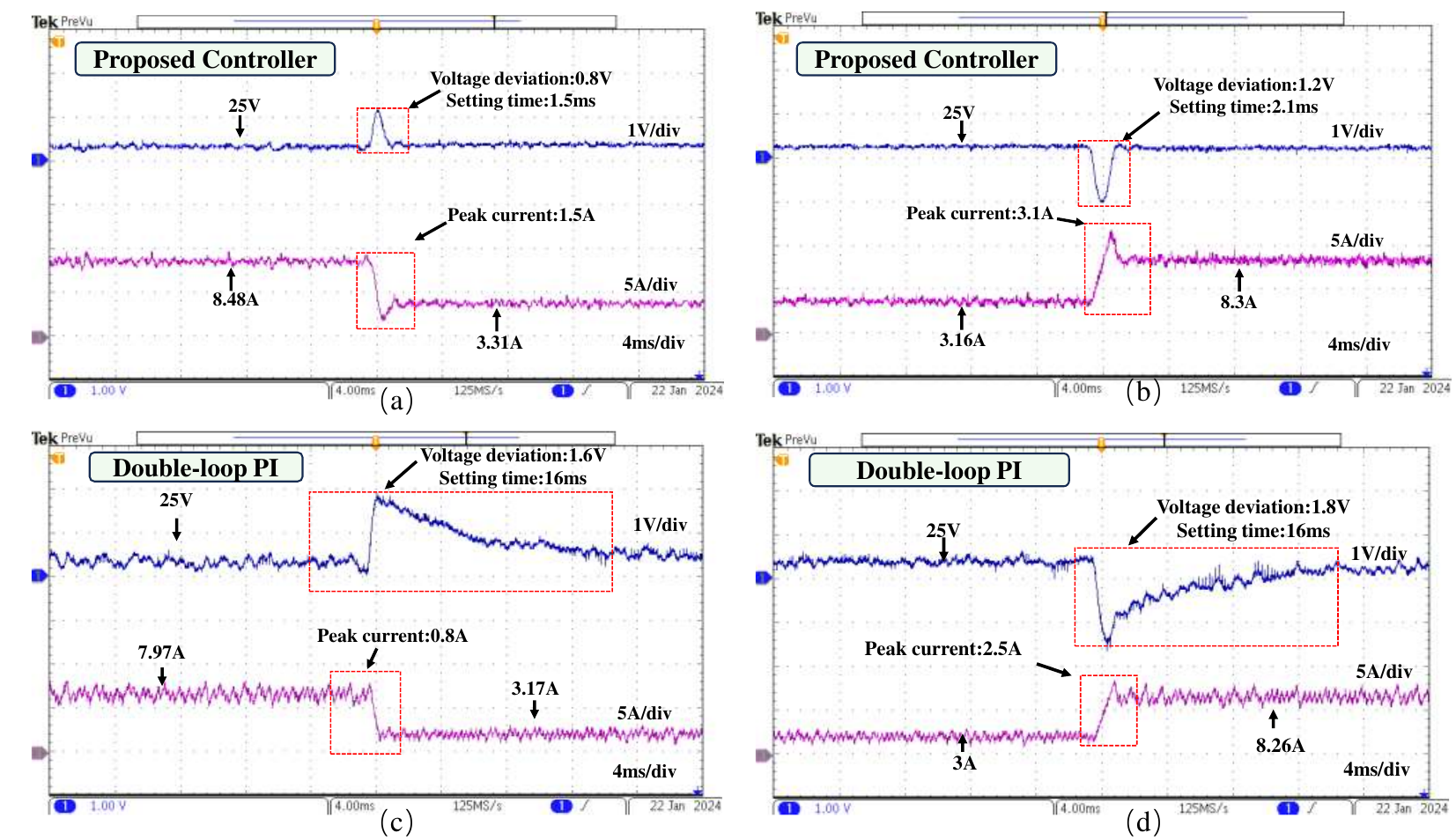}}
\caption{Experimental comparison results of case I.}
\label{exp1}
\end{figure}

Fig. \ref{exp1} illustrates the comparison between the PINN controller and the dual-loop PI controller. With $R_{imp}$ changing from 3$\Omega$ to 7$\Omega$, the proposed controller adjusts within 1.5ms, exhibiting a transient voltage deviation of 0.8V. Conversely, as $R_{imp}$ returns from 7$\Omega$ to 3$\Omega$, the adjustment time for the proposed controller's voltage regulation is 2.1ms, with a voltage deviation of 1.2V. $R_{imp}$ experiment results demonstrate the controller's capability to adaptively estimate unknown loads and maintain stable output voltage under resistive loads, which is challenging for conventional neural networks. 

\begin{itemize}
\item[$\bullet$] Case II. Comparisons analysis under $P_{CPL}$ conditions
\end{itemize} 

\begin{figure}[htbp]
\centerline{\includegraphics[width=1.1\linewidth]{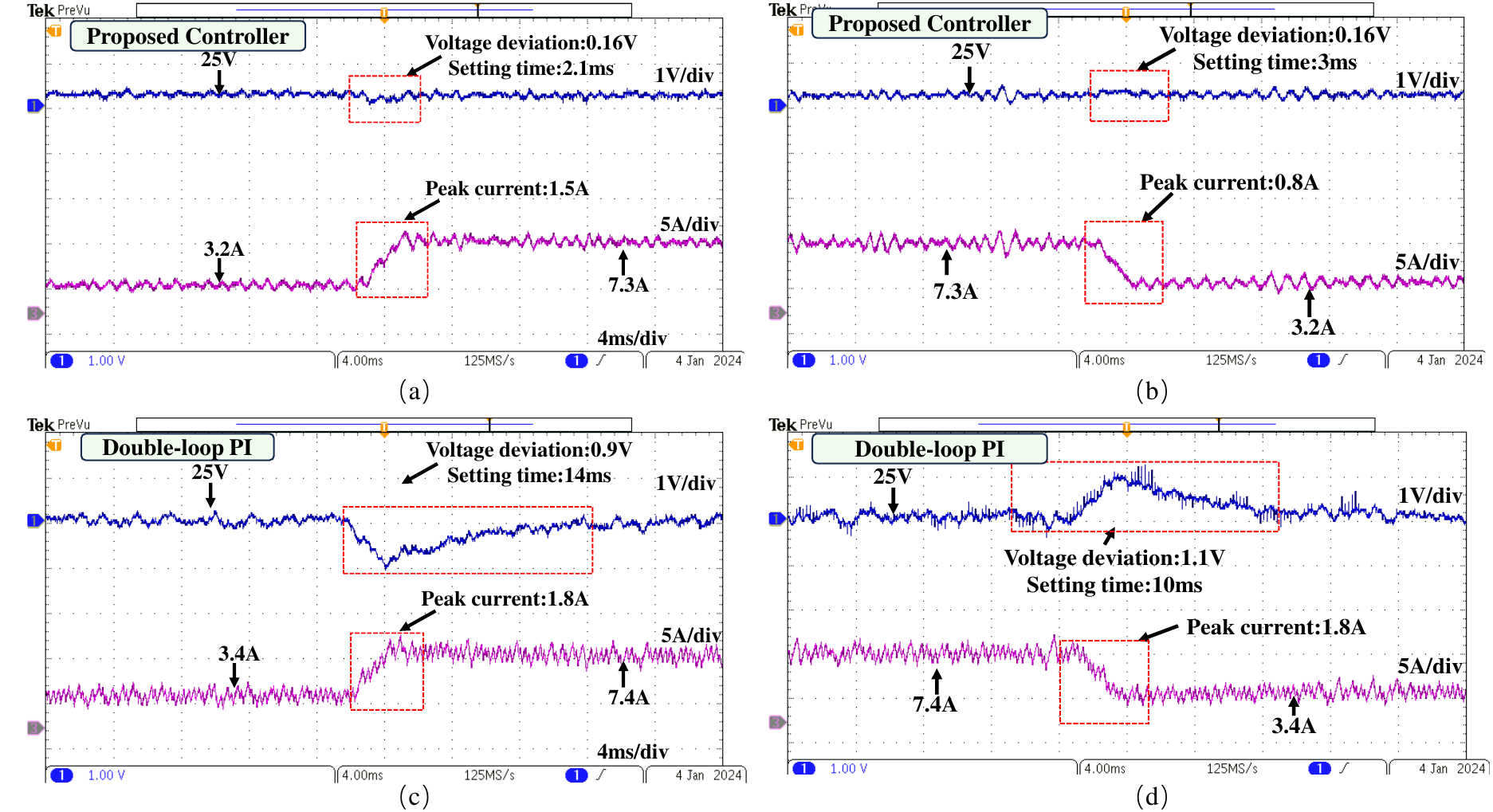}}
\caption{Experimental comparison results of case II.}
\label{exp2}
\end{figure}

When feeding a CPL load, the entire system can be considered as an inherent nonlinear system. As depicted in Fig. \ref{exp2}, when $P_{CPL}$ varies from 80W to 180W, the proposed controller adjusts within 2.1ms, exhibiting a transient voltage deviation of 0.16V.  Conversely, as $P_{CPL}$ returns from 180w to 80w, the adjustment time for the proposed controller's voltage regulation is 3ms, with a voltage deviation of 0.16V.
These outcomes demonstrate that the proposed controller could still maintain satisfactory output under CPL load conditions, proving the capability of the proposed controller to adapt to nonlinear disturbances. Furthermore, compared to the PI controller, it exhibits shorter settling times and smaller overshoots.

\section{Conclusion}

This paper presents a hybrid approach that combines model-driven and data-driven methodologies to address uncertainties in power electronics systems through the use of PINN. This method integrates uncertainty estimation and control laws based on physical principles with superior data-driven methods to develop a neural network. The effectiveness and generalization ability of the proposed method are verified through experiments. Although we only demonstrate the application of this method to a DC/DC buck converter in this paper, its potential for generalization and application across various power converter topologies with uncertain control systems is significant.  Therefore, in the future, this framework can be widely applied to different types of different power electronics topologies with uncertain control systems, and further studies will be conducted to address this intriguing issue. 

\bibliographystyle{ieeetr}
\bibliography{ref}

\vspace{12pt}

\end{document}